\documentclass[aps,prd,twocolumn,showpacs,superscriptaddress,groupedaddress,nofootinbib]{revtex4}  
\usepackage[dvipdfmx]{graphicx}  
\usepackage{dcolumn}   
\usepackage{bm}        
\usepackage{amssymb}   
\usepackage{amsmath}   
\usepackage{color}
\usepackage{ulem}

\begin{document}

\title{Comment on ``Relation between scattering amplitude and Bethe-Salpeter wave function in quantum field theory"}
\author{Sinya~Aoki}
 \affiliation{Center for Gravitational Physics, Yukawa Institute for Theoretical Physics, Kyoto University, Kyoto 606-8502, Japan}
\author{Takumi~Doi} 
\affiliation{RIKEN Nishina Center, RIKEN, Saitama 351-0198, Japan}
\affiliation{RIKEN Interdisciplinary Theoretical and Mathematical Sciences Program (iTHEMS), RIKEN, Wako 351-0198, Japan}
\author{Tetsuo~Hatsuda}
\affiliation{RIKEN Interdisciplinary Theoretical and Mathematical Sciences Program (iTHEMS), RIKEN, Wako 351-0198, Japan}
\affiliation{RIKEN Nishina Center, RIKEN, Saitama 351-0198, Japan}
\author{Noriyoshi~Ishii} \affiliation{Research Center for Nuclear Physics (RCNP), Osaka University, Osaka 567-0047, Japan}
\date{\today}

\begin{abstract}
We invalidate the  arguments given in  [T.~Yamazaki and Y.~Kuramashi, Phys.\ Rev.\ D {\bf 96}, 114511 (2017)]
over the HAL QCD method for hadron-hadron interactions on the lattice. 
 We also pose questions on the practical usefulness of the method  proposed in this reference.  
\end{abstract}

\preprint{YITP-17-125,~RIKEN-QHP-342}
\pacs{11.15.Ha,12.38.Gc}
\maketitle

In a recent article \cite{Yamazaki:2017gjl}, Yamazaki and Kuramashi present a theoretical analysis on the
``{\it relation between scattering amplitude and Bethe-Salpeter wave function in quantum field theory}".
 This subject  is important for hadron-hadron interactions in lattice QCD simulations and is closely 
  related to the physics of 
 multi-hadron systems  such as the exotic resonances and the atomic nuclei.

 In the present Comment, we show that the arguments in Ref.~\cite{Yamazaki:2017gjl}
  over the HAL QCD method  \cite{Aoki:2009ji,Aoki:2012bb,HALQCD:2012aa}  for hadron-hadron interactions
  can be invalidated  on the basis of the previously published  works by the present authors; in particular,   
Section 2, Section 3.1, Section 3.2, Appendix A and Appendix  B of  \cite{Aoki:2009ji}  as well as 
 Section~I of \cite{Aoki:2012bb}.

Let us consider the interaction between bosons with identical mass $m$ as analyzed in ~\cite{Yamazaki:2017gjl}.
The basic idea of the HAL QCD method is to define
 the energy-independent and non-local potential $U(\bold{r},\bold{r}')$ 
 \cite{Aoki:2009ji,Aoki:2012bb,HALQCD:2012aa} from the  Nambu-Bethe-Salpeter (NBS)
  wave function $\phi_{\bold{k}}(\bold{r})$ 
 below the inelastic threshold ($|\bold{k}| < {\rm k}_{\rm th}$) as 
  \begin{eqnarray}
(\nabla^2 + \bold{k}^2) \phi_{\bold{k}}(\bold{r}) & & =  m V (\bold{r}; \bold{k}) \phi_{\bold{k}}(\bold{r}) \nonumber \\ 
& & = m\int d^3 r' U(\bold{r},\bold{r}') \phi_{\bold{k}}(\bold{r}').
\label{eq:basic}
\end{eqnarray}

The primary  confusion of Ref.~\cite{Yamazaki:2017gjl} originates from a claim that
 $V(\bold{r};\bold{k})$ in Eq.(\ref{eq:basic})  is replaced by  $V(\bold{r};\bold{q})$ even for  $\bold{q} \neq \bold{k}$ 
in the HAL QCD method.
  Such a replacement however has {\it never} been introduced in the HAL QCD method.
 The correct mathematical relation between $U(\bold{r},\bold{r}')$ and $V (\bold{r}; \bold{k}) $ has been shown to be 
      (see  Eq.(2.3) in \cite{Aoki:2009ji} and   Eq.(7) in \cite{Aoki:2012bb})
      \footnote{For detailed analysis with inelastic channels, see Ref.~\cite{Aoki:2012bb}.}:
 \begin{eqnarray}
  U(\bold{r},\bold{r}') =  \sum_{|\bold{k}| < {\rm k}_{\rm th}} 
 \ V (\bold{r}; \bold{k}) \phi_{\bold{k}}(\bold{r}) \bar{\phi}_{\bold{k}}(\bold{r}'),
  \label{eq:U-V}
 \end{eqnarray}
where  
$\bar{\phi}_{\bold{k}}(\bold{r}')$ are dual basis functions associated with the non-orthonormal basis functions ${\phi}_{\bold{k}}(\bold{r}')$,
 with the property,  $ \int d^3 r \  \bar{\phi}_{\bold{k}}(\bold{r}) {\phi}_{\bold{q}}(\bold{r}) =\delta_{\bold{k},\bold{q}}$.
Eq.(\ref{eq:U-V}) represents a clear connection 
   between the $\bold{k}$-dependent local-potential $V (\bold{r}; \bold{k})$ and the $\bold{k}$-independent non-local potential $U(\bold{r},\bold{r}')$. 

In the practical applications of  the HAL QCD method \cite{Aoki:2009ji,HALQCD:2012aa}, 
 the expansion of $U(\bold{r},\bold{r}')$  in terms of its
  non-locality  is employed according to the well-known method of derivative expansion   \cite{TW,OM} 
  \begin{eqnarray}
  U(\bold{r},\bold{r}')  
   =  \sum_{\bf n} V_{\bf n}(\bold{r})  \nabla_{\bold{r}}^{\bf n} \delta^{(3)} (\bold{r}-\bold{r}'),
 \label{eq:V_i}
 \end{eqnarray}
 with $\nabla_{\bold{r}}^{\bf n} \equiv  \nabla_{r_x}^{n_x} \nabla_{r_y}^{n_y} \nabla_{r_z}^{n_z}$.
 (Here we have not assumed any spatial symmetry.)
  The $\nabla$-dependence
 is historically called the {\it velocity dependence}, since it is the origin of  the spin-orbit term,  the quadratic spin-orbit term, etc in the case of the 
  nuclear force.  (See the original articles, Refs.~\cite{OM,TW}, and Appendix B of Ref.~\cite{Aoki:2009ji}.) 
 Possible systematic errors associated with the truncation of this expansion can be  estimated through order-by-order analysis using  lattice QCD data \cite{Iritani:2017wvu,Kawai:2017goq,Iritani:2018zbt}. 


In  Ref.~\cite{Yamazaki:2017gjl},
there is  a confusion on the number of degrees of freedom.
  First of all, $U(\bold{r}, \bold{r}^\prime)$  is $\bold{k}$-independent from Eq.(\ref{eq:U-V}), so that
 $V_{\bf n}(\bold{r})$ is $\bold{k}$-independent from  Eq.(\ref{eq:V_i}).  Secondly,
  ${\bf n}$ (``$i$" in the notation of  Ref.~\cite{Yamazaki:2017gjl}) 
   is  a  three-dimensional vector, so that  there are sufficient  degrees of freedom in the right hand side of
    Eq.(\ref{eq:V_i}) to represent the    non-locality of  $U(\bold{r}, \bold{r}^\prime)$.
In fact, as  shown explicitly in Sec.II.D of Ref.~\cite{Kawai:2017goq} , the inversion of   Eq.(\ref{eq:V_i}) reads
 \begin{eqnarray}
 V_{\bf n}(\bold{r})  = \int d^3\bold{r}' \ U(\bold{r},\bold{r}') \frac{(\bold{r}' -\bold{r})^{\bold{n}}}{\bold{n}!} ,
 \end{eqnarray} 
 with $
 (\bold{r}' -\bold{r})^{\bold{n}}/{\bold{n}!}\equiv  \prod_{\ell=x,y,z} (r'-r)^{n_\ell}_\ell / {n_\ell!} $.
 This implies $V_{\bf n}(\bold{r})$ is manifestly ${\bf k}$-independent. 
  Here we note that the above integral 
  is expected to be  convergent below the inelastic threshold, 
 since there are no massless particles as asymptotic fields in QCD with 
 physical quark masses.

 In Ref.~\cite{Yamazaki:2017gjl}, there is  a statement that 
``{\it The problem in this expansion becomes manifest in the practical determination of $V_i(x)$. The simplest example is the determination of the leading term $V_0(x)$, which is approximated by $h(x;k)/\phi(x;k)$. We find that it should contain the contributions of $O(k^{2n})$ 
($n\ge 0)$ from the higher order terms of the velocity expansion in order properly describe the $k$ dependence of $h(x;k)$.}''.   
 This statement originates from a confusion of the
systematics  from the truncation of the derivative expansion with the ``$k$ dependence".
 To make this point clear, let us consider a hypothetical 
 non-local potential  $U^{(2)}$  which has only two terms in the right hand side of Eq.(\ref{eq:U-V}),
\begin{eqnarray}
 U^{(2)} (\bold{r},\bold{r}') = (V_0(\bold{r}) + V_2(\bold{r}) \nabla^2  ) \delta^{(3)}  (\bold{r}-\bold{r}').
\end{eqnarray}
Suppose we have two NBS wave functions at different momentums, 
$ \phi_{\bold{k}}(\bold{r})$ and $\phi_{\bold{k}^\prime}(\bold{r})$. 
Then, we can re-construct the potentials, $V_0$ and $V_2$, as
\begin{eqnarray}
    \left(  
    \begin{array}{c}
  V_0(\bold{r})  \\
  V_2(\bold{r})
   \end{array}
  \right)
  = \frac{1}{D}
    \left(  
    \begin{array}{cc}
  \nabla^2  \phi_{\bold{k}^\prime}(\bold{r})  & - \nabla^2  \phi_{\bold{k}}(\bold{r})     \\
  - \phi_{\bold{k}^\prime}(\bold{r}) &  \phi_{\bold{k}}(\bold{r})  
   \end{array}
 \right)
   \left(  
   \begin{array}{c}
  h(\bold{r}; \bold{k})  \\
  h(\bold{r}; \bold{k}^\prime) 
   \end{array}
   \right) ,
\label{eq:V02} 
\end{eqnarray}
with $D \equiv m( \phi_{\bold{k}}(\bold{r}) \nabla^2  \phi_{\bold{k}^\prime}(\bold{r}) 
- \phi_{\bold{k}^\prime}(\bold{r}) \nabla^2  \phi_{\bold{k}}(\bold{r})) $.
In this simple example, $V_{0}(\bold{r}) $ and $V_{2}(\bold{r}) $, which are 
$\bold{k}$-independent by definition,
reproduce the exact value of the phase shift $\delta(\bold{k})$ at all $\bold{k}$ below the inelastic threshold. 
In more realistic cases with higher order derivative terms in Eq.(\ref{eq:U-V}),  the phase shift $\delta(\bold{k})$ calculated from $V_{0,2}$ as constructed in Eqs.(\ref{eq:V02})  is exact 
 at  $\bold{q} = \bold{k}$ and $\bold{k}^\prime$, and is  only approximate at other $\bold{q}$.
  How accurate  at other $\bold{q}$
 can be checked if
more NBS wave functions for different momentums  are available. 
 In practice,  the time-dependent HAL QCD method based on  the Euclidean-time ($t$) dependence of the hadronic correlation function is a useful equivalent method to  treat those states with different momentums simultaneously, 
  as demonstrated  in \cite{Iritani:2017wvu}.

In Ref.~\cite{Yamazaki:2017gjl}, there is also a statement that 
``{\it Therefore, a smearing of the interpolating operator in the BS wave function gives a
different scattering amplitude from the one obtained from the fundamental relation, which
depends on the smearing function $s(x)$.}".
 As already shown explicitly in  Sec.II.D of \cite{Kawai:2017goq}, 
 this statement is mathematically incorrect.
 To clarify the source of this error, let us start with 
the well-known formula \cite{Lin:2001ek}
\begin{eqnarray}
 \phi_{\bf k}({\bf r}) &=& 
 C_{\phi}(k) e^{i {\bf k}\cdot {\bf r}} + \int \frac{d^3p}{(2\pi)^3} \frac{H_{\phi}(p;k)}{p^2-k^2-i\epsilon} 
 e^{i{\bf p}\cdot {\bf r}} ,
 \label{eq:asymptotic}
 \end{eqnarray}  
where the NBS wave function  $\phi_{\bf k}({\bf r})$  is defined by
\begin{eqnarray}
 \phi_{\bf k}({\bf r})  \equiv \langle 0 \vert {\pi}_1({\bf r}/2) {\pi}_2(-{\bf r/}2)\vert \hat{\pi}_1({\bf k}) \hat{\pi}_2(-{\bf k});{\rm in}\rangle ,
 \label{eq:NBS-non}
 \end{eqnarray} 
 with an interpolating operator of the $i$-th scalar particle ${\pi}_i$ ($i=1,2$), and 
 $C_{\phi}(k)$ being a normalization factor.
  The on-shell amplitude is related to the phase shift as  $H_{\phi}(k;k)/C_{\phi}(k)=4\pi e^{i\delta(k)} \sin\delta(k)/k$.

If we use the smeared NBS wave function $\tilde{\phi}_{\bf k}({\bf r})$
 with a smearing function, $s({\bf r})$, acting e.g. on one of the operators in  Eq.(\ref{eq:NBS-non}), 
 one easily obtains 
 \begin{eqnarray}
  \label{eq:HD}
 \frac{H_{\phi}(k;k)}{C_{\phi}(k)}=\frac{H_{\tilde{\phi}}(k;k)}{C_{\tilde{\phi}}(k)}
 =\frac{4\pi}{k} e^{i\delta(k)} \sin\delta(k), 
 \end{eqnarray}
   with $C_{\tilde{\phi}}(k)/C_{\phi}(k)= \int d^3r \ s({\bf r})  e^{-i{\bf k}\cdot {\bf r}}$ by a simple change of variable
(see Eq.(19) in  \cite{Kawai:2017goq}). Thus the smearing does not affect the observable
as long as the correct   normalization is taken into account.    Note that the situation is the same  even for composite sink-operators as long as they are 
    ``almost-local operator fields"  defined by Haag in  Ref.\cite{Haag:1958vt}.\footnote{See also a
     brief summary in Appendix A of \cite{Aoki:2009ji} on the Nishijima-Zimmermann-Haag reduction formula for the scattering of composite particles.}

Some additional comments are in order here.

Firstly, it is stated in Ref.~\cite{Yamazaki:2017gjl} that
``{\it Although we call it as the
effective potential here, the relation between this quantity and the potential in quantum
mechanics is not trivial; The former is the reduced BS wave function normalized by the
BS wave function being manifestly momentum dependent, while the latter is defined to
be momentum independent in principle. This is an essential difference between relativistic
quantum field theory and non-relativistic quantum mechanics.}"
This statement is inaccurate.  The difference between local (momentum dependent) and non-local (momentum independent)  potentials does not have  direct correspondence to the difference between 
 non-relativistic quantum mechanics and relativistic quantum field theory.
There are numerous examples where non-local (momentum independent) potentials 
appear  in non-relativistic quantum mechanics, especially for scattering problems of composite 
objects such as the atoms and atomic nuclei.

 Secondly, the so-called ``fundamental relation" (eq. (11)   in Ref.~\cite{Yamazaki:2017gjl})
\begin{eqnarray}
-\int d^3x \ h(x;k)e^{-i{\bf k}\cdot {\bf x}} &=& \frac{4\pi}{k}e^{i\delta(k)}\sin\delta(k)
\label{eq:fund}
\end{eqnarray} 
cannot provide an alternative method to determine the phase shift $\delta(k)$ in lattice QCD.
 Indeed, this equation is nothing but  Eq.(\ref{eq:HD}), so that the independent determinations of 
 the normalization $C_{\phi}(k)$ and  the momentum $k$ on the lattice are necessary.\footnote{
In contrast,  $U(\bold{r},\bold{r}')$ in Eq.(\ref{eq:basic}) does not have such normalization issue 
due to the cancellation of $C_{\phi}(k)$ in both sides \cite{Aoki:2009ji}.}
 If $k$ is known, however, 
  the phase shift $\delta(k)$ can be obtained  directly from the L\"{u}scher's finite volume formula \cite{Luscher:1990ux}
  without recourse to Eq.(\ref{eq:fund}).  Therefore, the above relation  gives at most a consistency check for the determination of $k$,  but does not provide  an alternative method to obtain $\delta(k)$. 
  Moreover, in the case that $k^2 < 0$ in the finite volume, the Fourier transformation of $h(x; k)$ must be analytically continued to the pure imaginary $k$ by the numerical integral, which may introduce uncertainties  due to
statistical errors of $h(x;k)$ at large $x$. These comments apply to
  a recent paper \cite{Namekawa:2017sxs} too.

\newpage

This work is supported in part 
by a priority issue (Elucidation of the fundamental laws and evolution of the universe) to be tackled by using Post ``K" Computer, 
and by Joint Institute for Computational Fundamental Science (JICFuS).
T.D. and T.H. were partially  supported by RIKEN iTHEMS Program.

\end{document}